\documentclass[acmsmall,natbib=false]{acmart}
\newcommand{\modelname}[0]{UPSR}
\usepackage[backend=biber]{biblatex}
\usepackage{hyperref}

\usepackage{graphicx} 
\usepackage{epstopdf} 
\usepackage{amsmath}
\usepackage{multirow}
\usepackage{booktabs}
\usepackage{array}
\usepackage{subfigure}
\usepackage{tabularx}

\usepackage{amssymb}
\usepackage{algorithmic}
\usepackage{algorithm}
\usepackage{verbatim}
\usepackage{bm}     
\usepackage{enumitem} 
\newcommand{\tabincell}[2]{\begin{tabular}{@{}#1@{}}#2\end{tabular}}
\usepackage{soul} 
\usepackage{color, xcolor} 
\AtBeginDocument{%
  \providecommand\BibTeX{{%
    \normalfont B\kern-0.5em{\scshape i\kern-0.25em b}\kern-0.8em\TeX}}}

\acmJournal{JACM}
\acmVolume{37}
\acmNumber{4}
\acmMonth{8}

\begin{document}

\title{Thoroughly Modeling Multi-domain Pre-trained Recommendation as Language}

\author{Zekai Qu}
\authornote{Both authors have equal contributions.}
\affiliation{\institution{China University of Geosciences Beijing}
\city{Beijing}
\country{China}}
\email{zekai_qu@163.com}

\author{Ruobing Xie}
\authornotemark[1]
\affiliation{\institution{Tencent Inc.}
\city{Beijing}
\country{China}}
\email{xrbsnowing@163.com}

\author{Chaojun Xiao}
\affiliation{\institution{Tsinghua University}
\city{Beijing}
\country{China}}
\email{xcjthu@gmail.com}

\author{Yuan Yao}
\affiliation{\institution{National University of Singapore}
\country{Singapore}}
\email{yaoyuanthu@163.com}

\author{Zhiyuan Liu}
\authornote{Corresponding author}
\affiliation{\institution{Tsinghua University}
\city{Beijing}
\country{China}}
\email{liuzy@tsinghua.edu.cn}

\author{Fengzong Lian}
\affiliation{\institution{Tencent Inc.}
\city{Shenzhen}
\country{China}}
\email{faxonlian@tencent.com}

\author{Zhanhui Kang}
\affiliation{\institution{Tencent Inc.}
\city{Shenzhen}
\country{China}}
\email{kegokang@tencent.com}

\author{Jie Zhou}
\affiliation{\institution{Tencent Inc.}
\city{Beijing}
\country{China}}
\email{withtomzhou@tencent.com}

\renewcommand{\shortauthors}{Trovato and Tobin, et al.}

\begin{abstract}
With the thriving of the pre-trained language model (PLM) widely verified in various NLP tasks, pioneer efforts attempt to explore the possible cooperation of the general textual information in PLM with the personalized behavioral information in user historical behavior sequences to enhance sequential recommendation (SR). However, despite the commonalities of input format and task goal, there are huge gaps between the behavioral and textual information, which obstruct thoroughly modeling SR as language modeling via PLM. To bridge the gap, we propose a novel unified pre-trained language model enhanced sequential recommendation (\modelname{}) that thoroughly transfers the next item prediction task to a text generation task, aiming to build a unified pre-trained recommendation model for multi-domain recommendation tasks. We formally design five key indicators, namely naturalness, domain consistency, informativeness, noise \& ambiguity, and text length, to guide the \emph{text}$\rightarrow$\emph{item} adaptation (selecting appropriate text to form the item textual representation) and \emph{behavior sequence}$\rightarrow$\emph{text sequence} adaptation (transferring the sequence of item textual representations into a text sequence) differently for pre-training and fine-tuning stages, which are essential but under-explored by previous works. In experiments, we conduct extensive evaluations on seven datasets with both supervised and zero-shot settings and achieve the overall best performance. Comprehensive model analyses also provide valuable insights for behavior modeling via PLM, shedding light on large pre-trained recommendation models. The source codes will be released in the future.
\end{abstract}

\ccsdesc[500]{Information systems~Language models}
\ccsdesc[300]{Information systems~Recommender systems}
\keywords{Recommendation, Language model, Pre-training}


\maketitle

\section{Introduction}
In our daily lives, recommender systems have been widely deployed for users to get information more efficiently. The personalized recommendation aims to provide appropriate items for users mainly based on their behaviors. Hence, sequential recommendation (SR) is broadly explored to make full use of historical behaviors, where users' interacted items (usually represented by item IDs) are viewed as the tokens in behavior sequence modeling. During training, the sequential behavioral patterns of different users are captured by SR models. Despite the wide usage of SR, most conventional SR models mainly focus on the behavioral information based on item ID via sequential modeling, and thus struggle with new items and domains with fewer interaction records \cite{SASRec}.


Recently, pre-trained language models (PLM)~\cite{GPT3,T5} have shown their magical power and are dominating in various NLP tasks. PLM can store common knowledge in the parameters with texts as bridges across different tasks. There are some commonalities between SR and PLM: (1) their inputs are certain token sequences (i.e., item or text sequences), and (2) their main training objective is the next token prediction task. Therefore, lots of classical sequence models in NLP~\cite{mikolov2010recurrent,Bert} have also been verified in SR~\cite{GRURec,BERT4Rec}. With the thriving of PLM, some pioneer works even begin to explore the possibility of directly modeling user behavior sequences in SR via PLM. These works intuitively achieve promising results in NLP-related tasks of recommendation (e.g., explainable recommendation~\cite{geng2022recommendation,gao2023chat}), while struggling with classical recommendation tasks~\cite{zhang2021language,liu2023chatgpt} due to the incompatibility between behavioral and textual modeling. Furthermore, Some works focus on using PLM as a modality information encoder to enhance item and user representation~\cite{UniRec,li2023text}, but still suffer from a suboptimal performance for the giant gap between the extracted representation and behavioral information.


We conclude four major differences between SR and PLM, which obstruct the direct usage of PLM in behavior modeling: (1) The central characteristic of recommendation is personalization, while PLMs are experts in common knowledge. (2) The tokens in PLM (texts) are often thousand-level and have concrete semantic meanings, while the tokens in SR (items) are more than million-level with obscure semantics. (3) Real-world SR suffers much more from extremely sparse user-item interactions in training. 
{(4) Compared to the diversified downstream tasks in PLM, pre-trained SR is more focused on dealing with the next item prediction task in different domains.}
These differences point to the intrinsic challenge of recommendation as language modeling: \emph{the huge gap between behavioral and textual information}. Better behavior-text mutual adaptations are required.


In this work, we attempt to take advantage of PLM in the behavior modeling of SR, bridging the gap between behavioral and textual information. We propose a novel \textbf{unified pre-trained language model enhanced sequential recommendation (\modelname{})}, which aims to build a unified behavior-tuned PLM as a pre-trained recommendation model to generate item texts for multi-domain SR, \emph{thoroughly} transferring the next item prediction task to a text generation task. Specifically, we spend effort on addressing three key challenges in behavior-text adaptations:
\begin{itemize}
    \item \textbf{\emph{How to build informative item textual representations from various practical item textual attributes?}} We first select some textual attributes of items to form a text sequence as the item's textual representation. However, practical textual attributes (e.g., title, category, brand, description) differ in the quality of representing items. We find that different textual attribute combinations as well as their orders have a great impact on the results of PLM-enhanced SR. To formalize this, we list five key indicators, namely \emph{Naturalness}, \emph{Domain consistency}, \emph{Informativeness}, \emph{Noise \& ambiguity}, and \emph{Text length}, to guide the \emph{text}$\rightarrow$\emph{item} adaptation.
    \item \textbf{\emph{How to construct appropriate text sequence from user historical behaviors for PLM to generate next items?}} To transfer historical behaviors to text sequences, we sequentially combine all items' textual representations in historical behaviors with a prefix user prompt to get a long text sequence as the input of PLM to generate the next item's texts. We also propose a masked item text prediction task to mingle both behavioral and textual information in PLM. The behavior-tuned PLM functions well for the \emph{behavior sequence}$\rightarrow$\emph{text sequence} adaptation.
    \item \textbf{\emph{How to design \modelname{}'s pre-training and fine-tuning for downstream multi-domain SR tasks?}} In training, \modelname{} first pre-trains the PLM via the user-item interactions in multiple pre-training domains to capture the general sequential behavioral patterns. Next, the behavior-tuned PLM is fine-tuned on the downstream target domain to learn domain-specific knowledge. To make full use of both general textual/behavioral knowledge in PLM/pre-training datasets and domain-specific behavioral preferences in the target dataset, we design different strategies in pre-training and fine-tuning.
\end{itemize}

Specifically, to adapt the item sequence to PLM, we select the item's textual attributes differently and set them in an appropriate order during the pre-training and fine-tuning stages to form a text sequence as the item's textual representation. We then combine all items' textual representations in user historical behaviors with a prefix user prompt to get a long text sequence as the input of PLM for text generation of the next item. 
In training, \modelname{} first pre-trains the PLM via the user-item interactions in multiple pre-training domains to capture the general user sequential behavioral patterns. Next, the unified behavior-tuned PLM is fine-tuned on the downstream target domain to learn domain-specific knowledge.


We conduct extensive experiments to evaluate our model on seven real-world datasets. \modelname{} achieves the overall best improvements on both pre-training and new domains.
We also verify that \modelname{}'s pre-training is beneficial even for zero-shot settings, and increasing the size of the pre-training dataset can further improve its performance. Besides, we conduct comprehensive analyses on different strategies of item textual representation construction and temporal robustness, aiming to reveal the insights of better using PLM for SR hidden in item representations.

The main contributions are concluded as follows:

\begin{itemize}
    \item We propose a novel unified pre-trained language model enhanced sequential recommendation, which thoroughly transfers the next item prediction task to a text generation task for recommendations in multiple domains.
    \item To bridge the gap between behavioral and textual information, we conduct in-depth explorations on the behavior-text adaptations (successes hidden in details), better activating PLM in behavior modeling and text generation of the next item.
    \item Extensive experiments on seven datasets have confirmed the effectiveness of our \modelname{} on both fine-tuning and zero-shot settings. Comprehensive model explorations and analyses on the gap between behavioral and textual information also bring in valuable insights to facilitate future research. 
\end{itemize}


\section{Related work}

In this section, we will briefly review previous works closely related to ours, including sequential recommendation and pre-training for recommendation.

\subsection{Sequential Recommendation}

Sequential recommendation (SR) aims to obtain the users' dynamic preferences based on their historical interactions. Early works on sequential recommendation usually capture sequential patterns through Markov chains (MCs). For instance, \citeauthor{FPMC} \cite{FPMC} fuses MF and MCs to model both general preferences and historical interactions. Besides the first-order MCs, there are also some methods adopting high-order MCs that take more previous items into account \cite{He_Kang_McAuley_2017,Ruining_Julian_2016}.
Later, RNNs such as Gated Recurrent Unit (GRU) \cite{GRU} are introduced to model user sequences, including session-based GRU (e.g., GRU4Rec \cite{GRURec}) and user-based GRU \cite{Donkers_Loepp_Ziegler_2017}. Besides, CNNs are also proved effective in modeling short-term sequential patterns using horizontal and vertical convolutional filters \cite{Tang_Wang_2018}. SG-GNN \cite{SR-GNN} obtains the local and global transitions by combining Graph Neural Networks (GNNs) with self-attention networks.

Recently, the success of sequence modeling in natural language processing (e.g., Transformers \cite{transformers}) draws broad attention to the possibility of using widely verified sequential models in NLP to model user behavior sequences. SASRec \cite{SASRec} employs a unidirectional Transformer to learn dynamic item transition patterns. BERT4Rec \cite{BERT4Rec} utilizes bidirectional Transformers with a cloze task as an additional training signal to better leverage the contexts for recommendation. LSSA \cite{LSSA} proposes a long and short-term self-attention network to capture both long-term preferences and short-term demands.
Recent efforts also bring in self-supervised learning (SSL) for SR \cite{qiu2022contrastive,chen2022intent,xie2022contrastive,wu2022selective,zhou2020s3}.
Some works explore cross-domain sequential recommendation \cite{ma2019pi,hao2021adversarial,cao2022contrastive}, while most of them are challenging to be smoothly transferred to new domains. To address this, besides adopting Transformers for behavior modeling, some works \cite{TransRec,UniRec} further use PLMs to fuse textual information into item representations for a more comprehensive understanding of item contents, functioning as the bridge across different datasets.

Different from these works, \modelname{} thoroughly transfers the behavioral sequence into texts, and directly uses a behavior-tuned PLM to generate item texts as results of SR.


\subsection{Pre-training in Recommendation}

Pre-training aims to learn general representations from large-scale data, thereby enhancing the performance on multiple downstream tasks. Recently, pre-trained language models \cite{Bert,GPT3,T5} and vision models \cite{vit,swim} have achieved great success for their superb performance and transferability.
In the field of recommender systems, pre-training has also gained great attention and achieved improvement. Early works attempt to leverage pre-trained models to fuse external modality information into user and item representations \cite{zeng2021knowledge,xiao2023uprec}. For example, some works acquire the side-information from the textual data through pre-trained word embeddings \cite{Gong_Zhang_2016,Zheng_Noroozi_Yu_2017,wu2021empowering}, and some possess it from knowledge graphs with pre-trained knowledge graph embeddings \cite{Zhang_Yuan_Lian_Xie_Ma_2016,Huang_Zhao_Dou_Wen_Chang_2018}. With the help of pre-trained knowledge, these methods could construct more expressive representations that are beneficial for recommendation. However, these works mainly use pre-trained models to extract external features separately and do not make full use of the power of PLMs.
Besides, some efforts of recommendation pre-training focus on learning generalized behavioral models and representations that could be smoothly transferred to downstream tasks. For instance, PeterRec \cite{peterrec} introduces the pre-training-and-fine-tuning paradigm to capture general sequential behavioral information and transfer learned user preferences for cold-start scenarios. \citeauthor{conure} \cite{conure} proposes the lifelong learning of recommendation based on one model, continuously updating user representations over time. ID-based pre-training models are also well explored \cite{wu2023attacking,wu2022personalized}. However, these methods are all based on the condition of certain shared IDs in both (pre)-training and test sets, which are limited in practical scenarios.

With the thriving of foundation models, some recent works propose to use them as item modality encoders (e.g., build item representations from their modality features based on PLMs rather than merely using the conventional ID embeddings), and design different pre-training tasks to learn more transferable representations across different recommendation scenarios. For example, IDA-SR \cite{IDA-SR} applies BERT to learn item representations from texts with three pre-training tasks derived from behavior sequences.  TransRec \cite{TransRec} utilizes both textual and visual item encoders for more transferable recommendations. Recently, some pioneer works explore directly using powerful LLMs for recommendation \cite{hou2023large,gao2023chat,wang2023zero}, while they cannot achieve comparable performance when interactions are sufficient since behavioral information is essential \cite{liu2023chatgpt,kang2023llms,hou2023large}.
P5 \cite{geng2022recommendation}, VIP5 \cite{geng2023vip5}, M6-Rec \cite{cui2022m6}, InstructRec \cite{zhang2023recommendation}, TallRec \cite{bao2023tallrec}, and AgentCF \cite{zhang2023agentcf} mainly focus on tuning PLMs via behavioral information for different types of downstream recommendation tasks in a single domain. UniSRec \cite{UniRec} aims to build a universal pre-training model for recommendations in multiple new domains. It adopts a fixed PLM with parameter whitening techniques to project multi-domain item embeddings into a shared space. RECFORMER\cite{li2023text} pre-trains a PLM on multiple domains considering both masked tokens and items and utilizes this model to encode both item textual representation and user sequence representation.

Different from these models, \modelname{} \emph{\textbf{thoroughly}} transforms SR into language modeling with a behavior-tuned PLM. It not only explicitly transfers the original historical behavior sequence into the text sequence as input, but also generates plain texts that represent the corresponding item as the result.

\begin{figure*}  
\centering  
\includegraphics[width=0.99\linewidth]{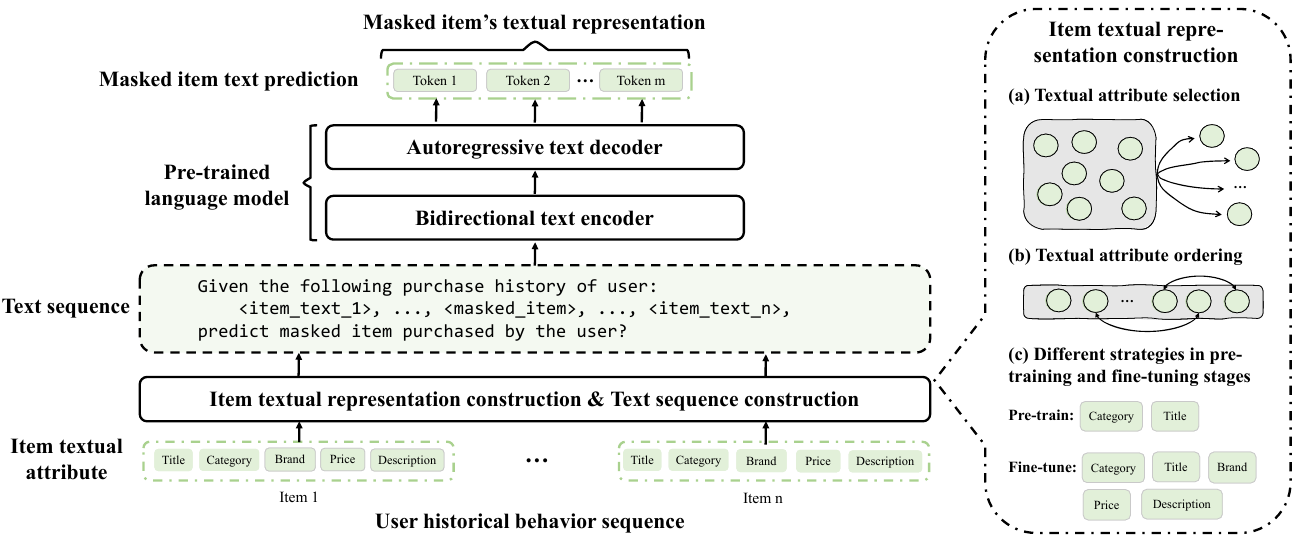}  
\caption{The overall framework of our proposed \modelname{} model. It transfers user historical behavior sequences into plain text sequences, and directly uses the PLM tuned on behavioral information of both pre-training and target recommendation datasets to understand user preferences and generate the next item's texts as the predicted result in downstream recommendation tasks.}
\label{fig:model}
\end{figure*}

\section{Methodology}
Our primary goal is to utilize PLMs for sequential recommendation (SR). In this section, we will first introduce the overall framework and task formulation of our \modelname{}. Second, we describe each component and its design principles in detail. Finally, we present the pre-training and fine-tuning process for PLMs in SR.



 
\subsection{Overall Framework}

\modelname{} aims to thoroughly transform SR into language modeling via behavior-tuned PLM. Precisely, we first need to represent the historical behavior sequences and items in textual formats. Then we treat the next item prediction task as a text generation task via PLM, which requires the model to directly generate the target item textual representation based on the text sequence of user historical behaviors. The overall framework of \modelname{} is in Figure~\ref{fig:model}.

Specifically, we denote $\mathcal{U}$ and $\mathcal{V}$ as the user and item sets. We define the historical behavior sequence as $S=\{v_1,v_2...v_n\}$, where each item $v_i$ is associated with several textual attributes, e.g., title, brand, etc.
First, at the item level, \modelname{} conducts a \emph{text$\rightarrow$item adaptation}: some textual attributes will be selected and orderly concatenated to obtain the item textual representation $R(v) =\{w_1,w_2...w_L\}$.
Next, at the sequence level, \modelname{} conducts a \emph{behavior sequence$\rightarrow$text sequence adaptation}: the behavior sequence will be converted into a text sequence $T = g(\{R(v_1),R(v_2)...R(v_n)\})$ with a prompt function $g(\cdot)$, which will be taken as the input of PLM for SR. 
Finally, the models are required to directly generate the recommended item textual representation based on $T$.

Since PLMs are pre-trained on language modeling tasks and have a large gap with SR, we conduct a behavior-enhanced pre-training for PLM via a novel masked item text prediction task with multi-domain historical interactions and then perform domain adaptation with fine-tuning. In this way, \modelname{} can effectively fuse (a) the common textual knowledge in the original PLMs, (b) the general behavioral pattern in behavior-enhanced pre-training, and (c) the domain-specific preferences in the target domain for better SR. 





\subsection{Key Indicators in Textual Representation}



Effectively textual representations of items and behavior sequences lie in the core of utilizing PLMs for behavior modeling. In real-world applications, each item will be associated with plenty of textual attributes. 
For example, an item in book recommendation should contain the book title, authors, the contents, the user comments, and so on. 
Ideally, item representations should include as many textual attributes as possible to enable them to (a) be comprehensive to reflect all types of user preferences, and (b) be distinguishable enough from other items. However, due to the limitation of PLMs in processing long sequences and the possible impact on the model's generality caused by the over-fitting issue, we adopt a strategy of selecting different textual attributes in different stages of \modelname{} to construct the most suitable item textual representations. We discuss five key indicators to guide our formalization as follows:

\begin{itemize}[leftmargin=*]
    \item \textbf{Naturalness}. It measures how closely an item's textual representation resembles natural language, which is the original processing target for PLMs. For example, the description of a book is usually a text that summarizes the content of the book in the form of natural language, which has a high naturalness. While the brand of cloth is usually composed of some unusual words (combinations) like Adidas, whose naturalness is extremely low. High naturalness will smooth the difference between item and text from the view of PLMs, better spurring PLM for downstream recommendation tasks in the NLP form and sharing generalized behavioral information between different domains. This indicator is mainly considered for selecting item textual attributes during the pre-training stage, which could be quantitatively reflected by inputting the item textual representation into PLMs and calculating its corresponding perplexity.
    
   \item \textbf{Domain Consistency}. 
   It indicates the semantic consistency of texts in item textual representation across different domains, and this indicator is mainly measured at the attribute level. Attributes having low domain consistency may have different semantics (e.g., ``Apple'' indicates different semantics in Electronics and Fruits fields) or reflect different user preferences (e.g., a brand may be sought after in one domain, but may not have a good reputation in another domain) in different domains, which may bring in conflicts in domain adaptations. Hence, item textual attributes with low domain consistency are suggested to be used in fine-tuning instead of pre-training. This indicator is hard to quantify and often empirically measured by manual inspection.
    \item \textbf{Informativeness}. It measures how much the item textual  representation contains item-related information. More textual attributes usually indicate higher informativeness.
    Generally, most PLMs are pre-trained on textual inputs that contain common semantic knowledge, which lacks the domain-specific knowledge for the downstream recommendation tasks. Adding additional information can effectively introduce item-related knowledge to models and achieve more accurate recommendations. However, excessive domain-specific information will lead to over-fitting and cause a decrease in model generalization. Therefore, the item textual representation should contain relatively less domain-specific information for pre-training to enhance our model's generalization ability. In contrast, in the fine-tuning stage, the item's textual representation should contain more information to be distinguishable from other items. We can calculate the information entropy of item textual representation as the quantitative metric of this indicator.
   \item \textbf{Noise \& Ambiguity}. It measures the possible noises and conflicts of texts in behavior-text transfer.
   Due to the uncontrollable data quality in real-world systems, some attributes may bring in irrelevant noises for prediction. For example, some product's titles may contain irrelevant stacked adjectives to attract users or cheat recommendation algorithms in E-commerce (e.g.,  oversized/slim-fitting/fashionable/elegant/cheap clothing). 
   In other cases, sometimes the same attribute value may have various meanings even inside a domain (e.g., books having the same title may not have similar contents, that is to say, some titles are not very related to their contents and user preferences). These will result in conflict and ambiguity in training SR based on the item's textual representations. Too noisy attributes may harm the performance, which should be carefully selected and pre-processed before using them.
    \item \textbf{Text Length}. It indicates the number of tokens of an item's textual representation. Although current LLMs (e.g., Longformer \cite{longformer}, LongNet \cite{longnet}) could model extreme long tokens as inputs, considering the possible long historical behavior sequence, too long texts are not that welcomed as the input of \modelname{}.
\end{itemize}
Ideally, the item textual representation  should be like natural language (to better extract knowledge from PLM and transfer general behavioral information between different domains), domain-consistent (to be smoothly adapted to other domains), informative (to accurately represent items and distinct them from other items), less noisy and ambiguous (for more accurate training and inference), and not too long (for PLM modeling). The goal in the behavior-text mutual adaptation is to make the semantic similarity of two item textual representations match their actual behavioral similarity.
\begin{table*}[!htbp]
\caption{Analysis of typical item textual attributes via five key indicators in Amazon datasets.}
\scalebox{0.78}{
\begin{tabular}{c|l|ccccc}
\toprule
Usage & Attribute   & Naturalness & Domain Consistency & Informativeness & Noise \& Ambiguity & Text Length \\
\midrule
\multirow{2}{*}{\tabincell{c}{Pre-training \\ \& tuning}} & Title       & high                 & high                     & high                  & medium         & high      \\ 
& Category    & medium                 & high                     & medium                  & medium         & low      \\ 
\midrule
\multirow{3}{*}{\tabincell{c}{Only used \\ in tuning}} & Brand       & medium                 & medium                     & medium                  & low         & medium      \\ 
& Price       & low                 & medium                     & medium                  & very low         & low      \\ 
& Description & high                 & medium                     & high                  & high         & very high      \\ 
\midrule
\multirow{2}{*}{Discard} & Related     & low                 & low                     & medium                  & high         & very high      \\ 
& SalesRank   & low                 & low                     & low                  & high         & low     \\
\bottomrule
\end{tabular}
}
\label{tab:key_metric}
\end{table*}


\subsection{Textual Representation Construction}

\subsubsection{Item Textual Representation Construction}
Item textual representation construction involves selecting appropriate attributes and concatenating them in a certain order.

\noindent
\textbf{Textual Attribute Selection}.
Considering the five key indicators discussed in the previous section, we need to avoid using textual attributes that are too long and have a large amount of noise, and construct different item textual representations at different stages of training so that the model can learn the corresponding behavioral information it prefers (i.e., more generalized behavioral information for pre-training, more domain-specific behavioral information for fine-tune). 
Here, we conduct a comprehensive analysis of the item's textual attributes. Without loss of generality, we show the qualitative results of the Amazon dataset in Table~\ref{tab:key_metric}, whose textual attributes are typical in practice.
We excluded the related attribute for its too-long text length and high noise, and also excluded the salesrank attribute for its low domain consistency and high ambiguity. 

Besides, we adopt different attribute selection strategies for pre-training and fine-tuning:
\begin{itemize}[leftmargin=*]
    \item {In the pre-training stage, the item textual representation should capture universal knowledge for the multi-domain recommendation. Hence, we design it to: (a) be more similar to natural language to cooperate well with PLM and smoothly transfer the generalized knowledge to new domains, (b) be less noisy for more accurate pre-training, (c) have higher domain consistency, so that the model can learn more generalized behavioral information for diverse new domains, and (d) contain relatively lower informativeness, especially on domain-specific knowledge, therefore easing the over-fitting issue of the model to the pre-training domains, which may cause damage to generalization. In this case, we choose (1) \emph{category}, (2) \emph{title}, to generate the item textual representation in pre-training. }
    \item {In the fine-tuning stage, the item textual representation focuses more on encoding more item-related information, and should follow these additional suggestions: (a) include more domain-specific information for the target domain and reduce the requirement for naturalness, so that the textual representation could capture more comprehensive downstream knowledge for prediction, and (b) have as much detailed information as possible to record behavioral information and avoid conflicts. Therefore, we additionally introduce (3) \emph{brand}, (4) \emph{price}, and (5) \emph{description} (their combination is noted as \emph{domain property text (DPT)}), on the basis of the textual attributes used in pre-training, making the item textual representation more informative and be able to capture the domain-specific behavioral information.}
\end{itemize}
{More quantitative analysis and experiment result about the selection of textual attributes will be presented in }\ref{sec.construction}.



\noindent
\textbf{Attribute Ordering.}
Besides attribute selection, we also need to concatenate them in a certain order to form the item textual representation, which proved to have a significant impact on the results in our empirical study but was under-explored in previous works.

In general, we find that arranging textual attributes from coarse-grained/general to fine-grained/ domain-specific has the best results in all domains. It may be because this ordering strategy could well simulate the real-world decision progress, which is also more suitable for the generation of PLM and could reduce the impact of noise on recommendation accuracy. Based on this finding, we arrange category$\rightarrow$title for pre-training and category$\rightarrow$title $\rightarrow$DPT for fine-tuning, and order DPTs according to the same standard. The textual attribute orders and combination format of different stages in \modelname{} are shown below:
\begin{quote}
\textbf{Item textual representation in pre-training}: 
(category: <category\_text>) <title\_text>
\end{quote}
\begin{quote}
\textbf{Item textual representation in fine-tuning:} (category: <category\_text>) <title\_text> (brand: <brand\_text>) (price: <price\_text>) (description: <description\_text>)
\end{quote}
where <\emph{attribute}\_text> indicates the texts of the \emph{attribute}'s value.

\subsubsection{Text Sequence Construction}

Thereafter, we need to convert the user historical behavior sequence into plain texts (i.e., \emph{text sequence}) as the input of PLM. The text sequence contains the following two parts: (a) a textual representation of the user's historical behavior sequence, and (b) an appropriate prompt to fully exert the capability of PLM for next item prediction. Specifically, we adopt the following prompt to build our text sequence:
\begin{quote}
\textbf{Text sequence}: Given the following purchase history of user: <item\_text\_1>, ..., <item\_text\_n>, <masked\_item>, predict masked item purchased by the user?
\end{quote}

Notably, prompt design is flexible and could be utilized to inject additional knowledge or personalized information to enhance model performance. For example, we can directly concatenate user profiles (e.g., age, gender, location under user approvals) in the prefix, and PLMs can build associations between user profiles and behavioral preferences. Moreover, the texts in the prompt may also impact the performance in different domains. We leave knowledge injection and detailed designs in prompts for future work. 





\subsection{Model Training and Inference}

In this section, we will illustrate how to pre-train, tune, and evaluate \modelname{} on multi-domain SR in the form of plain text. In this paper, we utilize two widely-used PLMs T5~\cite{T5} and FLAN-T5~\cite{FLAN-T5} as our PLM backbone.

\subsubsection{Model Training}

The central task of SR is next item prediction \cite{SASRec}. Recently, inspired by the great success of the masked token prediction (i.e., cloze) task in PLM \cite{Bert}, some SR models also adopt similar masked item prediction tasks \cite{BERT4Rec,UPRec,zhou2020s3} and achieve good results. Following these works, we propose a novel \textbf{masked item text prediction (MITP)} task for training, which could be viewed as a mixed task of masked item prediction in \cite{BERT4Rec} and masked token prediction in \cite{T5}.
Specifically, we randomly mask a proportion of items in the user historical behavior sequence (i.e., replace their corresponding <item\_text\_i> in the text sequence with sentinel token <extra\_id\_i>, where i is used to distinguish different sentinel tokens with a range from 0 to 99),
The model is then required to predict the masked items' texts in the original text sequence based on their context in both directions.

Although the MITP task can well learn the textual sequence modeling on behavior sequence, there exists an objective gap between MITP and the central next item prediction task, where sequential recommendation focuses more on recommending the next-1 items while MITP focuses on recommending items that are masked in the current context. 
To better align our model with the sequential recommendation, we randomly choose some samples that only mask the last item in users' text sequences during training. During both the pre-training stage and fine-tuning stage, we apply the same training task that mixes the masked (random) item text prediction and the masked (last) item text prediction by a ratio of nine to one, which obtains the best performances in the experiments.




Afterward, we feed the masked text sequence $T = [t_1,...t_n]$ as input into T5's text encoder and obtain its contextual representation. The decoder then attends to both encoder output and its previously generated tokens $y_{<j} = [y_1,...y_{j-1}]$ to predict the probability distribution of the next token, i.e., $P_{\theta}(y_j|T,y_{1:j-1})$. In training, \modelname{} optimizes its parameters $\theta$ by minimizing the mean cross-entropy loss of label tokens y in an autoregressive manner, which is formalized as follows:
\begin{equation}
    \mathcal{L}=-\frac{1}{|y|}\sum^{|y|}_{j=1}logP_{\theta}(y_j|T,y_{1:j-1}),
\end{equation}
where y is the concatenation of the masked items' texts to be predicted, and $y_j$ is the $j_{th}$ token in y.

\subsubsection{Model Inference}


As described above, when using \modelname{} for sequential recommendation, we mask the last item in the user sequence with a sentinel token <extra\_id\_0>, and then feed the masked text sequence into the model's encoder and obtain the user representation.

Next, for the candidate items set C =$\{c_1,...c_{m+1}\}$, where only $c_1$ is positive, we sequentially calculate the perplexity of the textual representation of each item based on the user representation, and recommend an item with the lowest perplexity. The calculation of the perplexity can be written as follows:
\begin{equation}
    PP(c_i) = P(w_1,w_2,...,w_N)^{-\frac{1}{N}},
\end{equation}
where N is the number of tokens of $c_i$'s textual representation, and $P(w_1,w_2,...,w_N)$ represents the probability that the model predicts this item textual representation.

Through the model training and inference, our \modelname{} successfully converts the behavioral sequence modeling in conventional sequential recommendation to the textual sequence modeling in NLP, taking advantage of both general knowledge in PLM and domain-specific behavioral preferences in user-item interactions.

\subsection{Connections and Differences with Other Pre-trained Recommendation Models}

Recently, there has been an emergence of pre-trained language model (or LLM) enhanced recommendations, aiming to take advantage of the magic power of LLMs in the field of recommender systems. We introduce the connections and differences of our \modelname{} with other representative types of pre-trained recommendation models to clarify our novelty and contributions:

\begin{itemize}
    \item PLM for prompt-based recommendations (e.g., ChatRec \cite{gao2023chat}). {This type of PLM-based recommendation evaluates the power of directly using PLMs/LLMs for sequential modeling in recommendation with manually designed prompts. Without behavior-aware tuning from pre-training and downstream recommendation domains, these models are good at zero-shot settings but unsatisfactory when items have a few interactions.}
    \item PLM with instruction-tuning (e.g., TallRec \cite{bao2023tallrec} and InstructRec \cite{zhang2023recommendation}). {These PLM-based models are trained mainly on (single-domain) user-item interactions or user instructions and thus have decent performance. However, they are still trained under the classical ranking losses, rather than directly predicting the next item as text generation used in our \modelname{}.}
    \item PLM for multi-task recommendations (e.g., M6-Rec \cite{cui2022m6} and P5 \cite{geng2022recommendation}). {It mainly focuses on handling different downstream tasks (e.g., SR, rating, explanation generation) in a single domain, rather than doing multi-domain recommendations like most recent PLM-based recommendation models, (e.g., UniSRec, RECFORMER, and our model). Besides, the ID information used in P5's items makes it challenging to directly use P5 for MDR.}
    \item PLM as item encoder (e.g., UniSRec \cite{UniRec}). {This type aims to build a universal pre-training model for SR in multiple new domains with the help of the modality information as a bridge. Different from other PLM models, it only uses PLM to build item embeddings from the item's modality information and still relies on conventional sequential models (e.g., SASRec) for behavior sequence modeling.}
    \item PLM as item and sequence encoder (e.g., RECFORMER \cite{li2023text}). {It converts items and behavior sequences into texts and adopts PLM to encode the text sequence. However, it still uses the conventional supervised objective for next item prediction, which is different from our text generation objective for the next item.}
\end{itemize}

{To the best of our knowledge, \modelname{} is the first to thoroughly transform multi-domain SR into language modeling via a behavior-tuned PLM in an end-to-end manner (for both input and output). We also conduct comprehensive analyses on different model settings (e.g., dataset/PLM sizes), tasks (e.g., zero-shot settings), and behavior-text adaptation strategies in building item textual representations.}

\section{EXPERIMENTS}
In this work, we aim to answer the following  questions:
\textbf{(RQ1)}: Can \modelname{} perform well in different downstream domains?
\textbf{(RQ2)}: Can the pre-trained \modelname{}  function in zero-shot setting?
\textbf{(RQ3)}: Is \modelname{} more suitable for (re)ranking?
\textbf{(RQ4)}: How to choose appropriate item textual attributes and orders in both pre-training and fine-tuning?
\textbf{(RQ5)}: How is the temporal robustness and generalization ability of \modelname{}?
\textbf{(RQ6)}: How is the performance of \modelname{} with different PLM model sizes and pre-training dataset sizes?


\subsection{Experimental Settings}

\textbf{Datasets.}
We conduct experiments on seven real-world datasets from the Amazon review datasets~\cite{amazon}. Without loss of generality, we use four datasets for pre-training and three datasets as new domains for tuning. (1) \textbf{Pre-training datasets}: we select widely-used \emph{Books}, \emph{Movies and TV}, \emph{Sports and Outdoors}, \emph{Clothing Shoes and Jewelry} as the pre-training domains. (2) \textbf{New datasets}: we select \emph{Arts, Crafts and Sewing}, \emph{Musical Instruments}, and \emph{Pantry} as the new domains for separate downstream tuning and evaluation, which are relatively sparser and has shorter behavior lengths. 
Following~\cite{SASRec,BERT4Rec}, for the pre-training dataset, we filter out users and items with less $5$ interactions and adopt their leave-one-out setting for building validation and test sets.

For new datasets, we just filter out the users with less than 4 interaction records to simulate a cold start scenario. Besides, for a fair comparison, we remove all valid/text instances that have untrained items (since most SR models cannot deal with them). Then we group the interaction records by users and sort them according to the timestamps. 
For each item textual representation, we truncate it with a maximum length of $40$, and for the whole user sequence, we truncate it with $512$ tokens for PLM. We do not align users having the same user ID across domains (very few) either for a more general and challenging setting. The detailed statistics are in Table \ref{statistic}.

\begin{table}[t]
\small
    \caption{Statistics of seven pre-training and new domain datasets in Amazon.}
    \label{statistic}
    \centering
    \begin{tabular}{l|rrrr}
        \toprule
        \textbf{Dataset} & \# \textbf{User} & \# \textbf{Item} & \# \textbf{Interactions} & \textbf{Avg.len}  \\
        \midrule
        \textbf{Pre-training} & 402,979 & 930,518 & 3,547,017 &8.8  \\
        - Books & 197,891 & 504,085 & 1,990,164 & 10.1  \\
        - Clothing & 135,041 & 294,788 & 1,004,679 & 7.4  \\
        - Sports & 39,477 & 87,235 & 262,998 & 6.7  \\
        - Movies & 30,570 & 44,410 & 289,176 & 9.5  \\
        \midrule
        \textbf{Arts} & 131,149 & 138,116 & 718,628 & 5.5   \\
        \textbf{Instruments} & 62,691 & 53,899 & 403,135 & 6.4  \\
        \textbf{Pantry} & 22,601 & 8,249 & 179,735 & 8.0  \\
        \bottomrule
    \end{tabular}
\end{table}


\noindent
\textbf{Parameter settings and metrics.}
We implement \modelname{} by PyTorch and Transformer package~\cite{transformer_package} and use T5-base and FLAN-T5 as our backbone PLM, which consists of a single Transformer of $12$ layers, $12$ attention heads and hidden state dimension $d=768$.
For tokenization, we directly use the SentencePiece~\cite{sentence_piece} tokenizer with a vocabulary size of $32,128$. In pre-training, we pre-train PLM for $20$ epochs using AdamW with $\lambda=3e^{-4}$ and the batch size set as $128$. In fine-tuning, we use the same learning rate with a batch size of $64$, using early stop with NDCG@10 to prevent overfitting. For both pre-training and fine-tuning, we set the mask ratio of the item as $0.1$. 
For each item textual representation, we truncate it with a maximum length of $40$, and for the whole text sequence, we truncate it with $512$ tokens for inputting into the model. 
We follow the classical setting~\cite{SASRec,BERT4Rec} and randomly sample $100$ negative items for evaluation. We adopt two widely-used metrics HR@k and NDCG@k with $k = 1,5,10$.

\noindent
\textbf{Competitors.}
To verify the effectiveness of our proposed approach, we compare it with some competitive (PLM-enhanced) SR baselines:
\begin{itemize}
    \item \textbf{SASRec}~\cite{SASRec}, which utilizes self-attention to model historical behavior sequence. It is a widely used and strong SR baseline in the industry.
    \item \textbf{BERT4Rec}~\cite{BERT4Rec}, which implements a bidirectional self-attention network with the cloze task for sequential modeling.
    \item \textbf{UniSRec}~\cite{UniRec}, which is the SOTA pre-trained model that uses PLM for the multi-domain recommendation. It first encodes item texts to embeddings via a fixed PLM, and then pre-trains and tunes a behavioral sequential model (SASRec) via behavioral information based on these item embeddings with domain adaptation modules.
\end{itemize}
For all of the hyper-parameters and model settings in the above models, we either follow the suggestion from the original paper or tune on the validation sets for fair comparisons. 
We do not compare with models like P5 \cite{geng2022recommendation} or TallRec \cite{bao2023tallrec} for they are mainly designed for single-domain tasks. We do not compare with models such as ChatRec \cite{gao2023chat} either since previous efforts have verified that the fixed LLMs are not suitable for being directly adopted to the recommendation. 
For UniSRec, we pre-train and fine-tune it on the same datasets of \modelname{} (utilizing the same textual attributes as our \modelname{} to encode the item representation, which has comparable performance compared to its original version).


\begin{table*}[!hbpt]
\small
\caption{Results of models on both pre-training and new domains (\%). \modelname{}~achieves the overall best performance (deviation about $\pm 0.3\%$). Note that the results of \emph{\textbf{new domains}} are more concerned when we evaluate pre-trained recommendation models. Our \modelname{}~uses one pre-trained PLM for all evaluations of pre-training and new domains, and thus is hard to achieve the best performance on all pre-training domains (we find that using pre-trained models with different training steps can further improve the performance in these domains).}
\label{performance}
\scalebox{0.85}{
\begin{tabular}{c|c|l|ccp{1.3cm}<{\centering}|ccr}
\toprule
Domain &  Dataset & Metric & BERT4Rec & SASRec & UniSRec & \modelname{} (T5) & \modelname{} (FLAN-T5) & Improv. \\
\midrule
\multirow{15}{*}{\tabincell{c}{New \\ domains}}

& \multirow{5}{*}{Arts}& NDCG@1        & 32.21               & {36.92}                                   & 36.81             & \underline{40.88}     & \textbf{42.79}           & \textbf{+15.89\%}        \\
&                             & NDCG@5      & 42.10               & {49.46}                                   & 48.67           & \underline{51.81}    & \textbf{53.70}          &    \textbf{+8.57\%}     \\
&                             & NDCG@10     & 45.41               & {52.62}                                  &  52.05         & \underline{54.84}      &\textbf{56.73}          &      \textbf{+7.39\%}   \\
&                             & HR@5        & 51.30          & {60.64}                                   &  59.36            & \underline{61.69}    &\textbf{63.70}           & \textbf{+5.05\%}        \\
&                             & HR@10       & 61.55               & {70.42}                                  & 69.84          &  \underline{71.03}      & \textbf{73.07}       &  \textbf{+3.76\%}       \\
\cline{2-9}
    & \multirow{5}{*}{Instruments} & NDCG@1        & 36.55               & {38.79}                             & 37.80            &   \underline{42.17}     & \textbf{43.74}         &      \textbf{+12.76\%}   \\
 &                            & NDCG@5      & 46.20               & {50.95}                              &  48.93        &  \underline{52.42}       & \textbf{54.50}        &     \textbf{+6.97\%}    \\
 &                            & NDCG@10     & 49.23               & {54.05}                              &  52.23          &  \underline{55.37}    & \textbf{57.41}           &     \textbf{+6.22\%}    \\
  &                            & HR@5        & 55.18               &  {61.77}                              & 59.01          &   \underline{61.91}       &\textbf{64.29}        &   \textbf{+4.08\%}      \\
 &                            & HR@10       & 64.45               & \underline{71.38}                             &  69.20          &   71.02  & \textbf{73.26}            & \textbf{+2.63\%}        \\
\cline{2-9}
&\multirow{5}{*}{Pantry}      & NDCG@1   &13.36   &15.22  &13.62     & \underline{18.32} & \textbf{18.72}  & \textbf{+23.00\%}\\
&                             & NDCG@5 &21.21    &24.76 &21.07       &\underline{26.72}  & \textbf{27.53} &\textbf{+11.19\%}\\
&                             & NDCG@10  &24.82   &28.63 & 24.69      &\underline{30.08} & \textbf{30.98}  & \textbf{+8.21\%}\\
&                             & HR@5    &28.84   &{33.97} &28.22       & \underline{34.77} & \textbf{35.85} & \textbf{+5.53\%} \\
&                             & HR@10   &40.08    &\underline{46.06} &39.51      &45.19  &\textbf{46.57} & \textbf{+1.11\%}\\
\midrule
\multirow{20}{*}{\tabincell{c}{Pre-training \\ domains}}
& \multirow{5}{*}{Books}
& NDCG@1              & 32.10                & {38.34}                             & 28.23       & \underline{39.82}   &\textbf{41.76}    & \textbf{+8.92\%}     \\
&                & NDCG@5     & 42.68              & {50.45}                                                    & 42.06        & \underline{50.68}    & \textbf{52.87}   & \textbf{+4.80\%}     \\
&                & NDCG@10    & 45.66              & 53.32                                        & 46.28                 & \underline{53.38}  & \textbf{55.49}     &  \textbf{+4.07\%}    \\
&                & HR@5       & 52.50              & \underline{61.32}
                 & 54.79                   & 60.73   & \textbf{62.64} & \textbf{+2.15\%}       \\
&                & HR@10      & 62.09              & \underline{69.73}                                           & 67.83                  & 69.05   & \textbf{70.72} &  \textbf{+1.42\%}       \\
\cline{2-9}
&\multirow{5}{*}{Clothing}    & NDCG@1               & 23.98                &  26.54                            & {28.60}        & \underline{31.79} &\textbf{32.63}      &  \textbf{+14.09\%}   \\
    &                             & NDCG@5     & 33.22               & 37.18                                    & {37.54}            &  \underline{41.42}   & \textbf{42.54}            &    \textbf{+13.32\%}    \\
&                             & NDCG@10    & 37.09               & 40.55                                     &  {41.08}         & \underline{44.50}    & \textbf{45.81}          &      \textbf{+11.51\%}   \\
&                & HR@5       & 42.30             &  {46.94}                & 46.07             &  \underline{50.47}       & \textbf{51.87}       &    \textbf{+10.50\%}     \\
&                             & HR@10      & 54.48               & {57.37}                                     & 57.05          & \underline{60.62}   & \textbf{61.99}             &     \textbf{+8.05\%}    \\
\cline{2-9}
&\multirow{5}{*}{Sports}      & NDCG@1        & 18.97              & 22.06                                    & {25.13}             &    \underline{26.66}   &\textbf{28.34}          &        \textbf{+12.77\%}  \\
&                             & NDCG@5      & 25.93              & 31.57                                   &  35.40           &   \underline{35.52}  & \textbf{37.50}            &        \textbf{+5.93\%} \\
&                             & NDCG@10     & 29.01              & 34.70                                     &  \underline{39.34}           &   38.58 & \textbf{40.59}           &    \textbf{+3.18\%}    \\
&                             & HR@5        & 33.00              & 40.64                                   &\underline{45.09}             &    44.10   & \textbf{46.14}         &        \textbf{+2.33\%} \\
&                             & HR@10       & 42.31              & 50.96                                    & \textbf{57.30}            &     54.30  & \underline{55.92}         &        - \\
\cline{2-9}
&\multirow{5}{*}{Movies}      & NDCG@1        & 37.68              & {40.28}                                     & 34.37            &  \underline{41.35}   & \textbf{43.43}            &      \textbf{+7.82\%}  \\
&                             & NDCG@5      & 48.55             & \underline{53.46}                                     &  44.96          &  52.40   &\textbf{54.74}            &       \textbf{+2.40\%}  \\
&                             & NDCG@10     & 51.72              & \underline{56.60}                                     & 48.32           & 55.61    & \textbf{57.69}            &      \textbf{+1.93\%}   \\
&                             & HR@5        & 58.63              & \textbf{{65.19}}                                    & 54.68          &   62.49   & \underline{64.85}         &       -  \\
&                             & HR@10       & 68.26              & \textbf{74.61}                                      & 65.09          & 72.22   & \underline{74.10}             &      -   \\
\bottomrule
\end{tabular}
}
\end{table*}

\subsection{Main Results (RQ1)}
\label{sec.main}

We compare \modelname{} with baselines on both new and pre-training datasets (new domains are more concerned for pre-trained models \cite{UniRec}). For a fair comparison, we remove all valid/test instances that have untrained items {(id-based method could not deal with this setting)}. The results are in Table \ref{performance}. We have:
\begin{itemize}
    \item \modelname{} achieves the best overall performance in most datasets, especially for three new domains with $12.76\%$-$23.00\%$ relative NDCG@1 improvements over the best baselines. It indicates that \modelname{} is capable of (a) effectively bridging the gap and transferring behavior sequences into text sequences, (b) flexibly extracting informative knowledge from PLM to capture user preferences from the transferred text sequences. \modelname{} outperforms baselines by a large margin especially in rank-related metrics such as NDCG@k, which implies that \modelname{} is more suitable for reranking. We also find that \modelname{} performs better than all baselines on all domains and metrics (including pre-training domains) as a reranking model (see Sec. \ref{sec.reranking}). {Besides, compared with the \modelname{} based on T5, the FLAN-T5 based \modelname{} achieves further improved performance, which has the same parameters  as the former, but induces the instruct tuning to  further enhance the capability of the language model. It indicates that with more advanced PLM, the recommendation performance of  our proposed framework could be further improved.}
    \item \modelname{} outperforms all classical SR models almost on all domains, which indicates that the textual information and the sequence modeling ability of PLM are beneficial for SR. Compared to UniSRec which also adopts PLM as a bridge for multi-domain recommendation, \modelname{} achieves better results on most datasets with more robust performance. {UniSRec directly encodes textual attributes into embeddings via a fixed PLM, which may bring information loss and intra-domain conflicts. The embeddings could not obtain the rich information originally contained in the text, thereby some fine-grained item attribute information will be neglected by the model. Besides, the domain-specific adapter does not have sufficient ability to align textual embeddings with behavioral patterns, so it encounters challenges when dealing with similar textual embeddings, especially for Books and Movies domains where titles have ambiguity. Our method could deal with the above problem by using a behavior-tuned PLM, which could use its sequence modeling ability to capture rich information in the text and distinguish an item's textual representation from others. }
    \item In general, \modelname{} has consistently better results on new and even pre-training domains. Considering the extremely challenging setting (\textbf{\emph{no explicitly aligned users, no overlapped items, one pre-trained model with the same parameters for all domains}}), the generalization ability of \modelname{} across different domains is satisfactory. We also observe that \modelname{} is better on sparser datasets with less popularity bias, where the supplement of textual information becomes more essential. Besides, \modelname{} has superior performance in zero-shot scenarios and reranking stages on all domains.
    \item To simulate the practical usage of pre-training models, we evaluate with the current challenging setting: adopting one unified pre-trained \modelname{} model for the tuning in all downstream domains. We find that the pre-training epoch and the weights of pre-training domains have large impacts especially for pre-training domains. The results in some domains (e.g., \emph{Sports}) will get further improvements with customized pre-training settings. However, UPSR is mainly designed for SR in new domains as conventional pre-trained models \cite{UniRec}. Hence, we conduct a balanced pre-training considering the generalization ability on different new domains. 
\end{itemize}
\begin{table*}[!hbpt]
\small
\centering
\caption{Results on zero-shot scenarios of new domains (\%).}
\label{zero-shot}
\begin{tabular}{c|p{3.6cm}|p{1.1cm}<{\centering}p{1.1cm}<{\centering}p{1.1cm}<{\centering}p{1.05cm}<{\centering}p{1.1cm}<{\centering}}
\toprule
Dataset & Model & NDCG@1  & NDCG@5 & NDCG@10 & HR@5 & HR@10\\
\midrule
\multirow{3}{*}{Pantry}
&UniSRec (Pre-training) & 05.74 &10.30 &12.77 &14.81 &22.52 \\
&\modelname{} (w/o Pre-training)  & 13.33 &17.25 &19.26 &21.03 &27.32  \\
&\modelname{} (Pre-training)    & \textbf{13.96} & \textbf{17.82} & \textbf{19.84} & \textbf{21.39} & \textbf{27.69}  \\
\midrule
\multirow{3}{*}{Arts}
&UniSRec (Pre-training) & 13.75 & 21.37 & 24.57 & 28.57 & \textbf{38.54} \\
&\modelname{} (w/o Pre-training) & \textbf{21.81} & \textbf{26.09} & \textbf{28.09}& \textbf{30.13} & 36.35  \\
&\modelname{} (Pre-training) & 20.27 & 24.73 & 26.74 & 28.96 & 35.23  \\
\midrule
\multirow{3}{*}{Instruments}
&UniSRec (Pre-training) &13.61 & 20.53 &23.03 & 26.86 &34.66 \\
&\modelname{} (w/o Pre-training) & 23.08 & 26.39 & 28.18 & 29.57 & 35.15  \\
&\modelname{} (Pre-training) & \textbf{23.42} & \textbf{27.57} & \textbf{29.41}& \textbf{31.48} & \textbf{37.21}  \\
\bottomrule
\end{tabular}
\end{table*}

\subsection{Results on Zero-shot Scenarios (RQ2)}

Conventional SR models cannot deal with the zero-shot scenario (new domains with no user-item interaction), which could be addressed by our text-enhanced \modelname{}. We evaluate three pre-training models on new domains without fine-tuning on the target domain: (a) \emph{UniSRec} uses its pre-trained model on pre-training datasets for evaluation. (b) \emph{\modelname{} (w/o pre-training)} directly uses the original T5 with our proposed text sequence (fine-tuning format) as input. (c) \emph{\modelname{} (pre-training)} adopts its pre-trained model as UniSRec. For a comprehensive comparison, all behaviors (except the first one) in new domains are viewed as test instances. In Table \ref{zero-shot} we have:
\begin{itemize}
    \item \modelname{} (pre-training) is capable of dealing with zero-shot scenarios without fine-tuning on the target domain, which achieves the overall best performance. It reconfirms the effectiveness of our proposed \emph{text}$\rightarrow$\emph{item} adaptation and \emph{behavior sequence}$\rightarrow$\emph{text sequence} adaptation in bridging the gap between behavioral and textual information, implying the practical usage of \modelname{} in cold-start domains. 
    \item \modelname{} (w/o pre-training) adopts PLM on this task with our proposed text sequences as the inputs. Surprisingly, it also achieves promising results (worse than the pre-trained version on Pantry and Instrument while better on Art, since the four pre-training domains have fewer supplements to Art). 
    \item As for UniSRec, it performs unsatisfactorily in zero-shot scenarios, since sufficient tuning of its adapters is essential for new domains. Both \modelname{} versions outperform UniSRec, which also verifies the effectiveness of our constructions of item textual representation and text sequence.
\end{itemize}


\subsection{Results on \modelname{} as a Reranking Module (RQ3)}
\label{sec.reranking}

\begin{figure}[htbp!]
    \centering
    \includegraphics[width=0.9\textwidth]{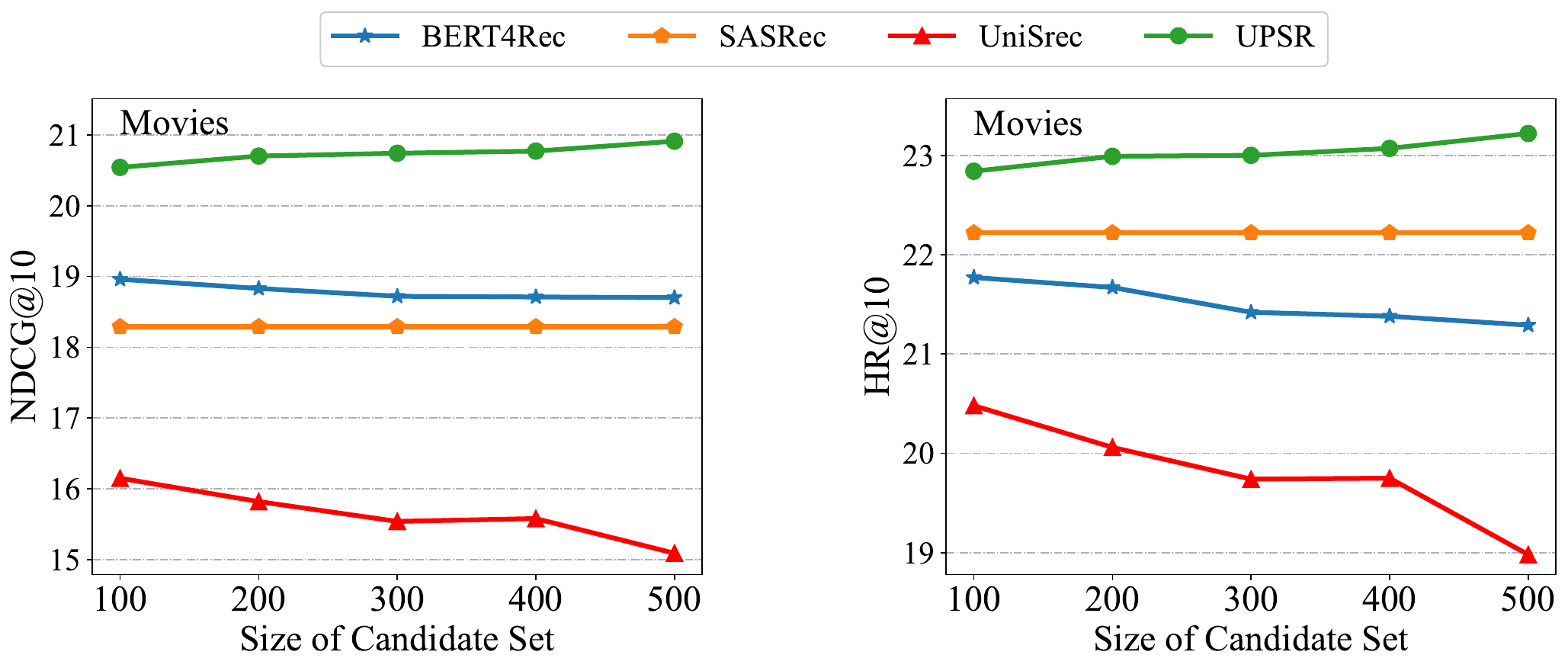}
    \includegraphics[width=0.9\textwidth]{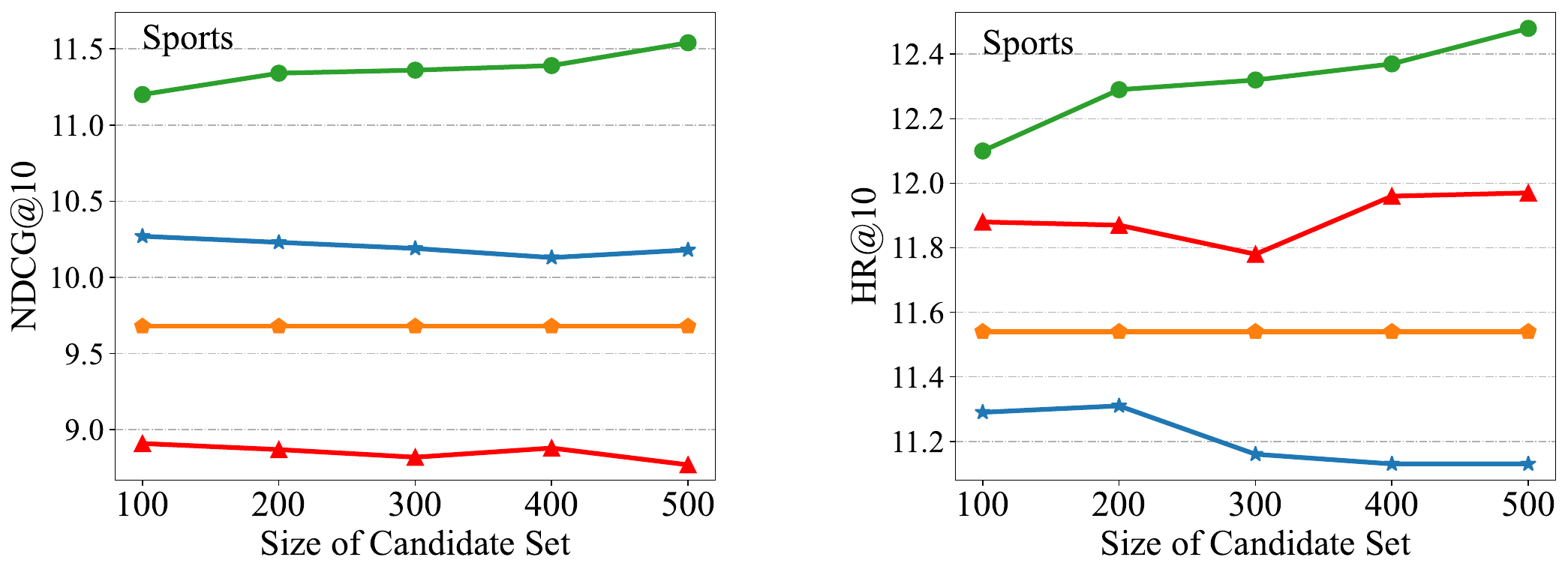}
    \caption{Results of \modelname{} and baselines as reranking module.}
    \label{rerank}
    
\end{figure}

In Table \ref{performance} and \ref{zero-shot}, we find that \modelname{} achieves larger improvements on NDCG@k, implying that \modelname{} is more suitable for distinguishing harder negative samples in the reranking stage widely existed in real-world systems.
Hence, to simulate the reranking stage, we adopt the widely-used SASRec as the preceding matching/ranking model to retrieve several hundred items from all items corpus, and then evaluate all models' reranking ability from these retrieved items (we choose T5 as the backbone of \modelname{} in this setting). From Figure \ref{rerank} we find that:

\begin{itemize}
    \item  \modelname{} significantly outperforms all baselines on all datasets and metrics in reranking with different candidate sizes (including underperforming Movies and Sports in Table \ref{performance}). It reconfirms the power of \modelname{} especially for distinguishing challenging negative items in practical reranking.
    \item The results of \modelname{} are monotonically increasing with the growth of candidate size, while other baselines are fluctuating or declining. It indicates that baselines suffer seriously from harder negative items, while our model is robust since it well fuses both behavioral and textual information, and thus could distinguish both types of negative items.
    \item Considering the advantage of \modelname{} in reranking and its efficiency, we believe that such PLM-enhanced recommendation should be deployed in reranking rather than matching in real-world systems.
\end{itemize}

\subsection{In-depth Analyses on Item Representation Construction from Texts (RQ4)}
\label{sec.construction}

Building appropriate item textual representations with Table \ref{tab:key_metric} is decisive in \modelname{}. In this section, we give comprehensive analyses of the details of attribute selection and ordering in pre-training and fine-tuning. T, C, D indicate title, category, and DPT, respectively.

\subsubsection{Which Item Textual Attributes Should We Select?}

To determine what textual attributes should be selected to construct the item textual representation during fine-tuning, we mainly focus on two aforementioned indicators, namely Informativeness and Ambiguity. Therefore, we evaluate three attribute combinations on \modelname{} (single), namely directly using the original T5 to fine-tune on one target domain, as: (a) title, (b) title+category, and (c) title+category+DPT, which have an increasing amount of information and a decreasing degree of ambiguity. 

From Table \ref{attribute selection} we can find that: both category and DPT are beneficial in \modelname{}, because they both bring extra information and reduce conflicts between textual representations of different items in the same domain. To be more specific, the category highlights the essential domain information to distinguish different domains' behavior sequences, while DPT aims to distinguish items in the same domain for better item representation learning, working as a coarse-grained semantic-related item ID. Therefore, the category brings significant improvement in both datasets, and for Sports where item titles are distinguishable enough, DPT does not bring in much additional improvement. While for Movies where item titles often have conflicts, DPT is essential to reduce the mismatching between semantic similarity and behavioral similarity, and thus improves the performance by a large margin. This result is consistent  with the Informativeness we calculated for different item textual representations. When the category was introduced, the Informativeness of textual representation improved greatly on both Sports (3.94 $\rightarrow$4.79) and Movies (2.77$\rightarrow$4.12), while DPT only brought a small improvement on sports (4.79$\rightarrow$4.83), but having a significant impact on movies (4.12$\rightarrow$4.68). It indicates that the indicators we propose are able to effectively reflect the impact of textual attributes on \modelname{}'s  performance, and we can utilize these indicators to guide us in selecting the components of item textual representation.



\begin{table}[!hbtp]
\small
\caption{Results of different item attribute combinations (\%).}
\label{attribute selection}
\begin{tabular}{c|l|p{1.25cm}<{\centering}p{1.28cm}<{\centering}p{1.28cm}<{\centering}}
\toprule
Dataset                 & Metric  & T  & T+C & T+C+D \\
\midrule
\multirow{5}{*}{Sports}  & NDCG@1   & 22.23  &  25.74       &  \textbf{26.00}           \\
                        & NDCG@5 & 30.70 &   34.91       & \textbf{35.13}    \\     
                        & NDCG@10 & 34.44 & 38.14         & \textbf{38.23}              \\
                          & HR@5 & 39.46 &   43.62       &  \textbf{43.76}   \\   
                        & HR@10   & 50.60 & 53.30         & \textbf{53.43}              \\
\midrule
\multirow{5}{*}{Movies}   
                        & NDCG@1   & 30.20 & 33.46         &   \textbf{40.53}           \\
                        & NDCG@5 & 40.01  & 45.19          & \textbf{51.56}    \\      
                        & NDCG@10 & 43.51 & 48.97         & \textbf{54.80}    \\      
                          & HR@5 &49.33  & 56.21         &  \textbf{61.96}   \\   
                        & HR@10   & 60.73 & 67.85         & \textbf{72.00}              \\
                        
\bottomrule
\end{tabular}
\end{table}


\subsubsection{How to Set Appropriate Order when Combining Selected Textual Attributes?}

Besides the elements, the order of attributes in item textual representation also matters. In Table \ref{attribute_order}, we evaluate three orders on \modelname{} (single) as (a) T$\rightarrow$C$\rightarrow$D, (b) D$\rightarrow$C$\rightarrow$T, (c) C$\rightarrow$T$\rightarrow$D.

We observe that: C$\rightarrow$T$\rightarrow$D achieves the best performance on both datasets. {This result is also consistent with our calculation of Naturalness for item textual representation with different attribute orders that the composition of C$\rightarrow$T$\rightarrow$D has the lowest perplexity, while the Informativeness of the three is roughly the same. Therefore, we assume that an appropriate order should arrange attributes from coarse (e.g., category) to fine (e.g., DPT), which is a more natural form for PLM to understand and learn preferences from. It could smooth the gap between item and text, making PLM easier to understand the context and generate reasonable recommendations. Moreover, the category functions well as a domain-specific delimiter of items (indicating the beginning and end of an item's text), making the text sequence more understandable for PLM.}



\begin{table}[!hbpt]
\small
\caption{Results of different textual attribute orders (\%).}
\begin{tabular}{c|l|ccc}
\toprule
Dataset                 & Metric    & T$\rightarrow$C$\rightarrow$D & D$\rightarrow$C$\rightarrow$T & C$\rightarrow$T$\rightarrow$D   \\
\midrule
\multirow{5}{*}{Sports} & NDCG@1  & 25.30     & 25.68        & \textbf{26.00} \\
                        & NDCG@5    & 34.09     & 34.58        & \textbf{35.13} \\
                        & NDCG@10   & 37.54     & 37.74        & \textbf{38.23}\\
                         & HR@5  & 42.57     & 43.34        & \textbf{43.76} \\
                        & HR@10 & 53.40     & 53.22        & \textbf{53.43} \\
\midrule
\multirow{5}{*}{Movies} & NDCG@1  & 40.45     & 37.53        & \textbf{40.53} \\
                        & NDCG@5    & \textbf{51.60}     & 49.37        & 51.56 \\
                        & NDCG@10   & 54.65     & 52.79        & \textbf{54.80}\\
                         & HR@5  & 61.78     & 60.37       & \textbf{61.96} \\
                        & HR@10 & 71.96     & 70.93        & \textbf{72.00} \\
\bottomrule
\end{tabular}
\label{attribute_order}
\end{table}

\subsubsection{Should We Select Different Textual Attributes in Pre-training and Fine-tuning?}

Besides, we find that the item attribute selection should be different in pre-training and fine-tuning, in that case, \modelname{} is able to capture more generalized behavioral patterns in the source domain and effectively adapt to the target domain. We report four strategies for the attribute selection in pre-training/fine-tuning: (a) T/T, (b) C+T/C+T, (c) C+T+D/C+T+D, (d) C+T/C+T+D. In Table \ref{attribute_difference}, we have the following findings from the main scenario:
\begin{itemize}
    \item In fine-tuning for a specific domain, both category and DPT (brand, price, description) are beneficial to \modelname{}'s performance since they could provide more detailed domain-specific information of items. It is natural that more item information is always welcomed in recommendations considering the serious sparsity issue.
    \item In pre-training for all domains, C+T and C+T+D achieve relatively comparable performance on Instruments and Pantry, while the former performs better on metric that is more related to the model's generalization (i.e., HR), and the latter performs better on metric related to model's accuracy (i.e., NDCG). Meanwhile, C+T achieves the overall best performance on Arts. We assume that this is because the Instruments and Pantry have a high correlation with the pre-training dataset, but the source domain has fewer supplements to Art. Therefore, C+T+D could utilize the highly related information to make more precise recommendations on Instruments and Pantry, while C+T enables \modelname{} to obtain more generalized knowledge and achieve better performance on less-related domains.
\end{itemize}

\begin{table}[!hbpt]
\small
\caption{Results of different item textual attribute selection strategies (pre-training/tuning) in the main scenario (\%).}
\begin{tabular}{c|l|p{2.0cm}<{\centering}p{2.0cm}<{\centering}p{2.15cm}<{\centering}p{2.15cm}<{\centering}}
\toprule
Dataset                  & Metric    & T/T  &{C+T/C+T}   & {C+T+D/C+T+D}     & {C+T/C+T+D}     \\
\midrule
\multirow{5}{*}{Instruments} 
                            &NDCG@1	&33.12 &39.43& \textbf{42.27}  &\underline{42.17}\\
                            &NDCG@5	&43.98 &49.45&\textbf{52.54}   &\underline{52.42}\\
                          & NDCG@10 &47.59 &52.70 & \textbf{55.50} & \underline{55.37}\\
                          &HR@5	&54.28 &58.70&\underline{61.84}  & \textbf{61.91}\\
                            &HR@10  &65.53 &68.74& \underline{70.96}  &\textbf{71.02}\\
\midrule
\multirow{5}{*}{Pantry} 
                            &NDCG@1	&13.46 &16.57 & \textbf{18.71}  &\underline{18.32}\\
                            &NDCG@5	&19.31 & 23.60 &\textbf{26.79}   &\underline{26.72}\\
                          & NDCG@10 &22.42  &26.70& \textbf{30.16} & \underline{30.08}\\
                          &HR@5	&24.98 &30.34&\underline{34.42}  & \textbf{34.77}\\
                            &HR@10  &34.68 & 39.99 & \underline{44.88}  &\textbf{45.19}\\
\midrule
\multirow{5}{*}{Arts} 
                            &NDCG@1	&31.80 &39.41 & \underline{40.75}  &\textbf{40.88}\\
                            &NDCG@5	&42.11 & 50.03 & \underline{51.65}   &\textbf{51.81}\\
                          & NDCG@10 &45.85  &53.17& \underline{54.70} & \textbf{54.84}\\
                          &HR@5	&51.83 &59.74&\underline{61.53}  & \textbf{61.69}\\
                            &HR@10  &63.41 & 69.47 & \underline{71.00}  &\textbf{71.03}\\
\bottomrule
\end{tabular}
\label{attribute_difference}
\end{table}

\begin{table}[!hbpt]
\small
\caption{Results of different item textual attribute selection strategies (pre-training/tuning) in the zero-shot scenario (\%).}
\begin{tabular}{c|l|p{2.0cm}<{\centering}p{2.0cm}<{\centering}p{2.15cm}<{\centering}p{2.15cm}<{\centering}}
\toprule
Dataset                  & Metric    & T/T  &{C+T/C+T}   & {C+T+D/C+T+D}     & {C+T/C+T+D}     \\
\midrule
\multirow{5}{*}{Instruments} 
                            &NDCG@1	&\textbf{24.22} &19.44&18.40   &\underline{23.42}\\
                            &NDCG@5	&\textbf{28.60} &23.01&21.27   &\underline{27.57}\\
                          & NDCG@10 &\textbf{30.83} &25.14 &23.29  & \underline{29.41}\\
                          &HR@5	&\textbf{32.88} &26.58&24.13  & \underline{31.48}\\
                            &HR@10  &\textbf{39.82} &33.24&30.47  &\underline{37.21}\\
\midrule
\multirow{5}{*}{Pantry} 
                            &NDCG@1	&\textbf{14.66} &13.48 & 11.54  &\underline{13.96}\\
                            &NDCG@5	&\textbf{18.73} & 17.19 &15.21   &\underline{17.83}\\
                          & NDCG@10 &\textbf{20.70}  &19.34& 16.93& \underline{19.84}
                          \\
                          &HR@5	&\textbf{22.53} &20.71&18.67  & \underline{21.39}\\
                            &HR@10  &\textbf{28.71} & 27.43 & 24.02  &\underline{27.69}\\
\midrule
\multirow{5}{*}{Arts} 
                            &NDCG@1	&\underline{19.70} &17.00 & 16.94  &\textbf{20.27}\\
                            &NDCG@5	&\underline{24.07} & 21.00 & 21.17   &\textbf{24.73}\\
                          & NDCG@10 &\underline{26.24}  &22.96& 23.17 & \textbf{26.74}\\
                          &HR@5	&\underline{28.28} &24.88&25.17  & \textbf{28.96}\\
                            &HR@10  &\underline{35.05} & 30.99 & 31.42  &\textbf{35.23}\\

\bottomrule
\end{tabular}
\label{attribute_difference_zero}
\end{table}


Furthermore, We also report these strategies for the attribute selection in zero-shot scenarios, from Table \ref{attribute_difference_zero}, we can find that:

\begin{itemize}
    \item T/T achieves the overall best performance in zero-shot scenarios, it is because this type of  item textual representation is closest to natural language and PLM could smoothly transfer generalized behavioral information between different domains.
    \item C+T/C+T+D achieves nearly comparable performance as T/T and even surpasses it on Arts, which verifies that our construction of item textual representation could effectively narrow the gap between text and behavior. 
    \item C+T+D/C+T+D gets the worst performance in this scenario, which indicates that utilizing C+T+D for pre-training might introduce too much domain-specific information and impair the generalization of \modelname{}. 
    \item C+T/C+T+D achieves a significant improvement in performance on all datasets compared to C+T/C+T in zero-shot scenarios since the former introduces more item-related information that helps  PLM to understand items.
\end{itemize}

Jointly considering all above analyses in both main and zero-shot scenarios, we finally adopt C$\rightarrow$T for pre-training and C$\rightarrow$T$\rightarrow$D for fine-tuning, so that our \modelname{} could acquire more generalized behavioral pattern from pre-training datasets and obtain sufficient domain-specific information to make precise recommendations on downstream domains. 
In practice, algorithm designers could choose their preferred textual attributes, orders, and strategies according to their specific demands with our findings in online systems.

\subsection{Analyses on Temporal Robustness (RQ5)}
\label{sec.robustness}

\begin{figure}[htbp!]
    \centering
    \includegraphics[width=0.45\textwidth]{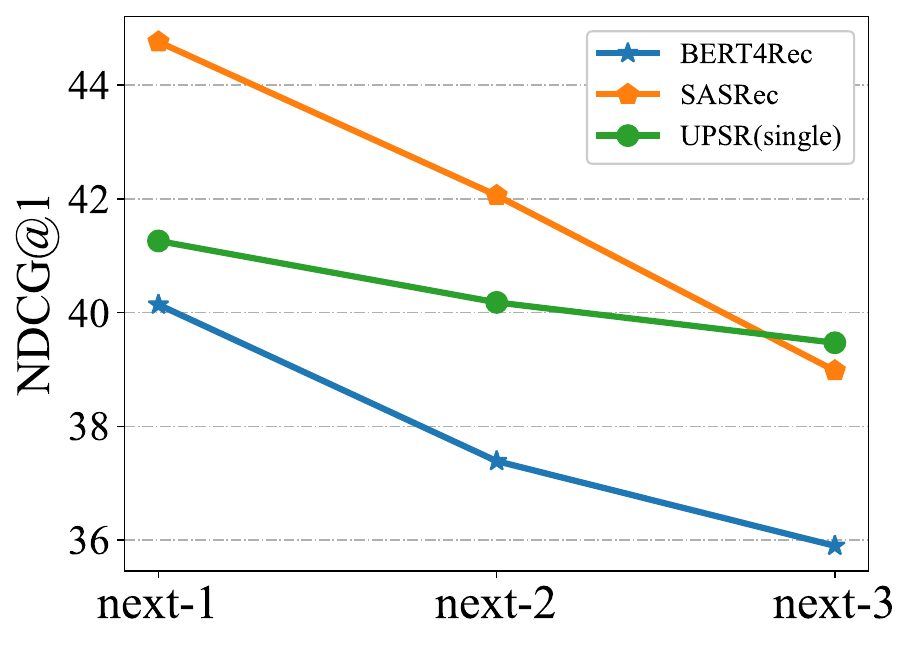}
    \includegraphics[width=0.45\textwidth]{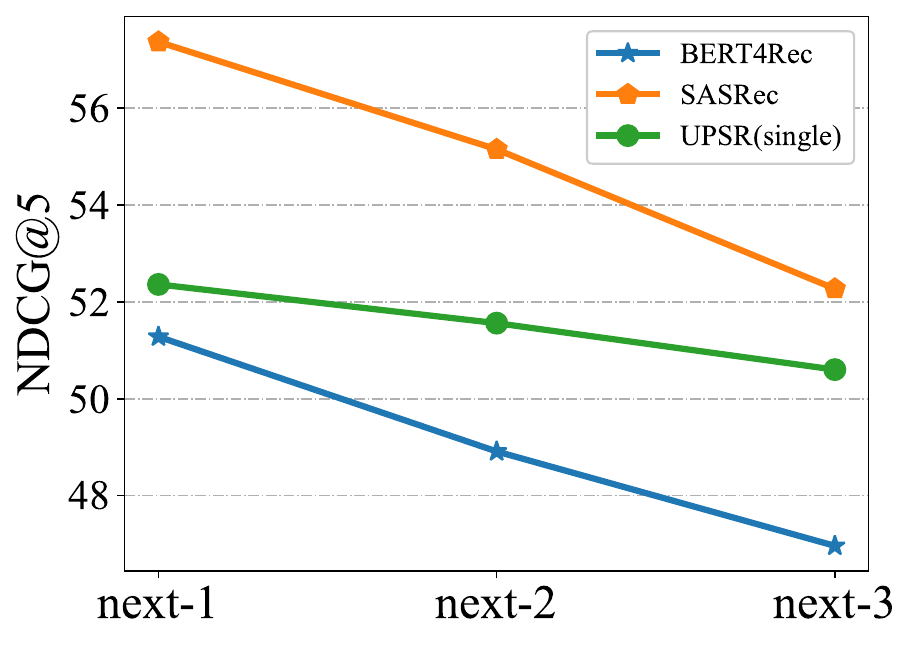}
    \caption{The results of temporal robustness on next-k prediction. We find that our \modelname{} is more robust in next-k prediction as k increases.}
    \label{robustness}
\end{figure}

During evaluation, we find that lots of conventional SR models have severe overfitting issues under the leave-one-out setting, since the last interacted item of a user in the train set is temporally close to the test instance. For example, our SASRec and BERT4Rec achieve their best performance with nearly $600$ epochs. However, practical recommendation models should deal with multiple consecutive recommendation requests of users after one full model training (e.g., daily update). In this subsection, we attempt to evaluate the temporal robustness of SR models on predicting next-k items to avoid overfitting on recent interactions, which is a more practical and challenging setting. Specifically, we leave the last three interacted items of each user as the test set on Movies, retrain all models, and report the results of next-1, next-2, and next-3  in Figure \ref{robustness}. Based on the figure, we draw the following observations:
\begin{itemize}
    \item All three models achieve their best results on the next-1 items and the worst results on the next-3 items. It implies that the current training objective and the leave-one-out  setting will inevitably result in serious overfitting issues, which focuses too much on next-1 clicks but neglects generalization on next-k clicks. 
    \item The NDCG@1 of BERT4Rec and SASRec drop sharply over time (nearly $10.6\%$ and $12.9\%$), while \modelname{} keeps relatively stable (nearly $4.3\%$). It is because \modelname{} could well balance both recent behavioral information and long-term textual information from user historical behaviors, which alleviates the overfitting issue and is more robust in practical recommender systems.
\end{itemize}


\subsection{Analyses on Model and Data Sizes (RQ6)}
\label{sec.analyses_size}


\subsubsection{Analysis on PLM Size}

In this subsection, we will discuss the influence of the PLM's model size on the final performance. For comparison, we replace T5-base with T5-small under the exact same setting on the Clothing and Sports datasets. The parameter numbers of T5-small and our backbone T5-base are $61M$ and $223M$, respectively. 
From Table \ref{tab:PLM_size}, we can observe that when the parameter size of PLM is reduced to a quarter of T5-base, the performance only drops slightly. {This implies that our \modelname{} might be able to be employed on other smaller PLM, retaining the performance of \modelname{} basically unchanged, which is more computationally friendly.} 
The efficiency is essential to practically deploy PLM-based models in online recommendations for efficiently dealing with billion-level real-world user requests. 
Moreover, since we have discovered that FLAN-T5 performs better than T5 from Table \ref{performance}, it is also worthy to evaluate with larger PLMs to explore whether there are also emergent abilities in recommendation with our framework in the future.



\begin{table}[!hbpt]
\small
\caption{Results of different PLM sizes (\%).}
\begin{tabular}{c|l|p{1.32cm}<{\centering}p{1.32cm}}
\toprule
Dataset & Metric    & T5-small     & T5-base    \\
\midrule
\multirow{5}{*}{Sports} 
                            &NDCG@1	&25.82 &\textbf{26.00}\\
                              &NDCG@5	&34.59 &\textbf{35.13}\\
                           &NDCG@10 &37.63  &\textbf{38.23}\\
                            &HR@5	&42.86 &\textbf{43.76}\\
                            &HR@10  &52.65 &\textbf{53.43}\\
                          
\midrule
\multirow{5}{*}{Clothing}   
                            &NDCG@1	 & 31.25  &\textbf{31.32}\\
                            &NDCG@5	 &40.77 &\textbf{41.09}\\
                           &NDCG@10 &43.71  &\textbf{44.28}\\
                            &HR@5	&49.62  &\textbf{50.20}\\
                            &HR@10  &58.70 &\textbf{60.36}\\
\bottomrule
\end{tabular}
\label{tab:PLM_size}
\end{table}

\subsubsection{Analysis on Pre-training dataset size}

We also investigate how the size of pre-training data affects the results. Precisely, we select more domains from Amazon to build an even larger pre-training dataset (nearly $36.7M$ interactions, $10$ times the original one). {Then we pre-train both T5 and FLAN-T5 with this larger pre-training dataset, and fine-tune them on Instruments and Pantry datasets under the exact same setting as the smaller one. Specifically, we compare the performance of three sizes of the pre-training dataset, including: ``none'', ``origin'' and ``large'', where ``none'' denotes the vanilla PLM without behavior-aware tuning, ``origin'' reflects the \modelname{} pre-trained on the small pre-training dataset, and ``large'' signifies the \modelname{} with the larger pre-training dataset.}

\begin{figure}[htbp!]
    \centering
    \includegraphics[width=0.87\textwidth]{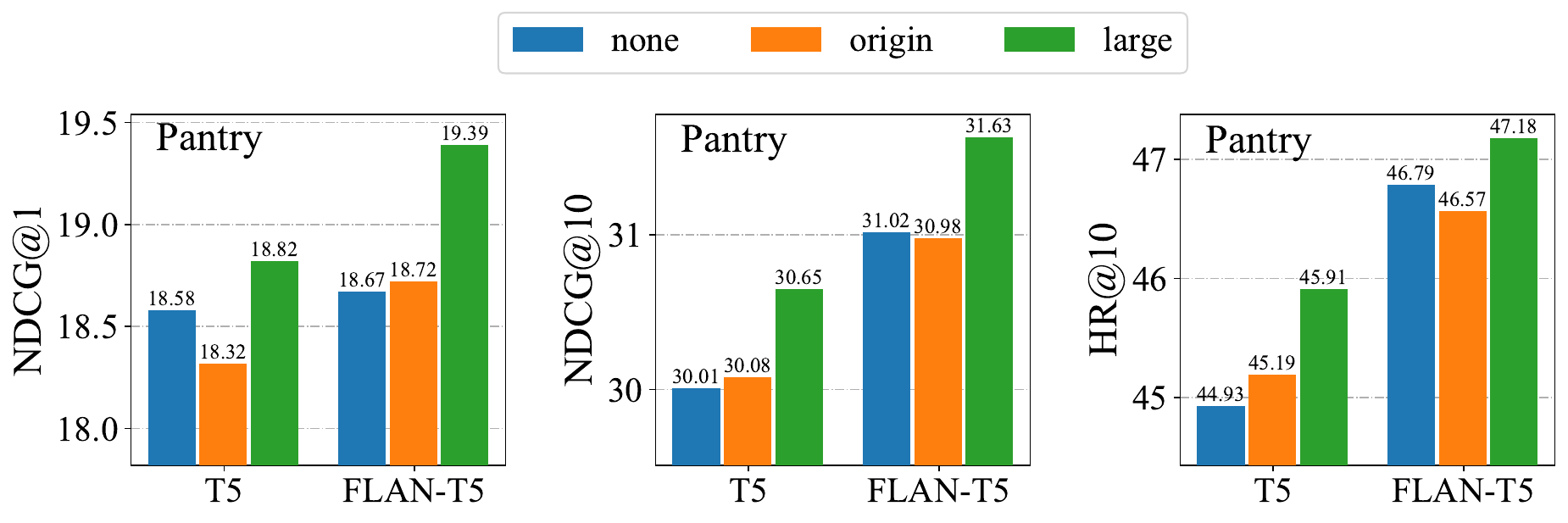}
    \includegraphics[width=0.87\textwidth]{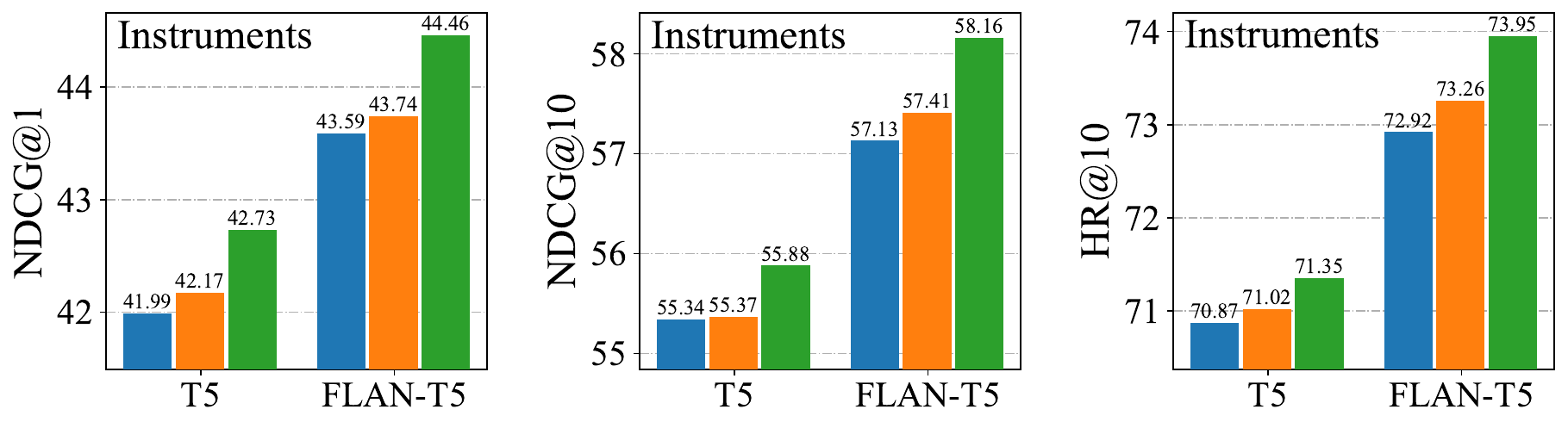}
    \caption{Results of different pre-training dataset sizes in the main fine-tuning scenario.}
    \label{bigger-pretrain}
\end{figure}
\begin{figure}[htbp!]
    \centering
    \includegraphics[width=0.87\textwidth]{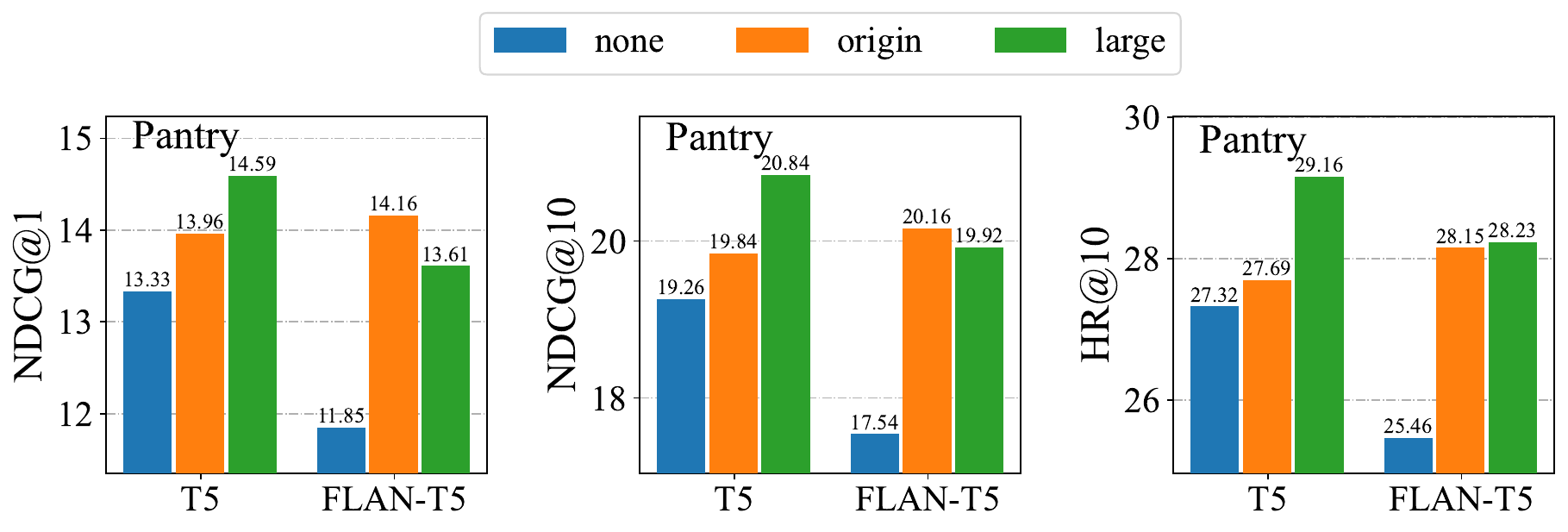}
    \includegraphics[width=0.87\textwidth]{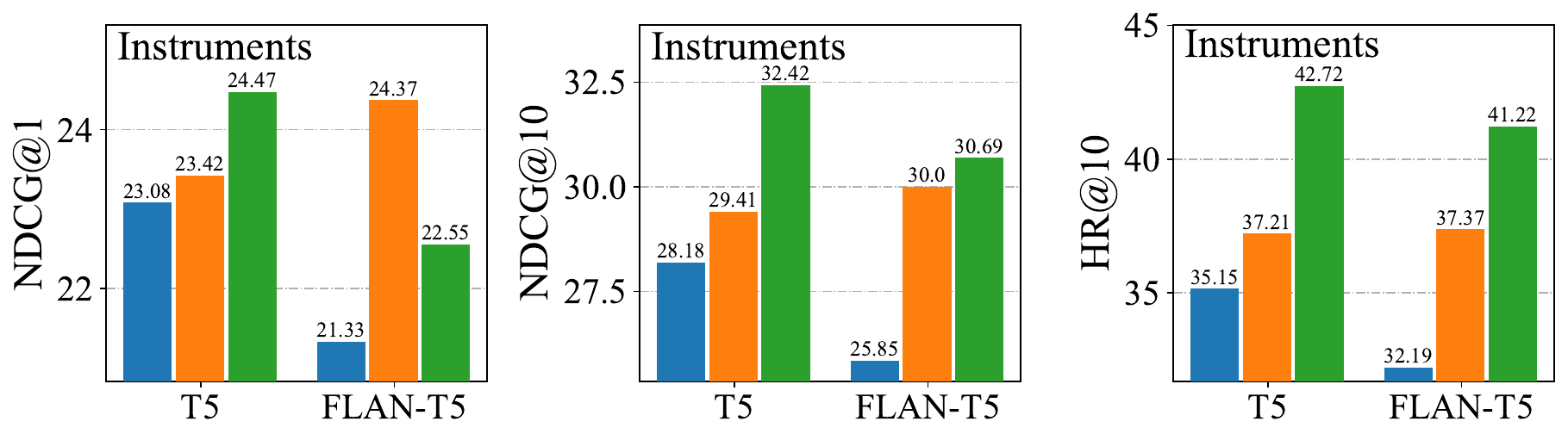}
    \caption{Results of different pre-training dataset sizes in the zero-shot scenario.}
    \label{bigger-pretrain-zeroshot}
\end{figure}

We summarize the results in Figure \ref{bigger-pretrain} and Figure \ref{bigger-pretrain-zeroshot}, where we have the following observations: 
\begin{itemize}
    \item As the size of the pre-training dataset increases, \modelname{} achieves even better results on both Pantry and Instruments for different PLM. 
    \item FLAN-T5 generally outperforms T5, especially with larger datasets in fine-tuning scenarios. This indicates that FLAN-T5 may have better learning capabilities  for leveraging the additional information during fine-tuning. 
    \item In zero-shot scenarios, the difference in performance between T5 and FLAN-T5 is more pronounced, the vanilla T5 is more powerful in making recommendations than vanilla FLAN-T5, and with the expansion of the pre-training dataset, the zero-shot capacity of T5 is continuously improving while FLAN-T5 only needs the smaller pre-training dataset to obtain a satisfactory performance, and will suffer from performance decline when expanding the dataset. This may be due to the instruction fine-tuning of FLAN-T5 on multiple tasks, which enables it to quickly adapt to a new type of task, but makes it have poor zero-shot ability on a completely unseen task.
\end{itemize}

In the future, it is worth evaluating with real-world-level larger pre-training datasets to explore whether our proposed model could handle practical SR tasks.

\section{Conclusion and Future Work}


In this work, we propose a novel unified pre-trained language model enhanced sequential recommendation for multi-domain SR, which thoroughly transfers the next item prediction task to a text generation task supported by a behavior-tuned PLM. Extensive experiments with detailed analyses verify the effectiveness and robustness of \modelname{}, shedding light on future large pre-trained recommendation models.

In the future, we will jointly consider more modality information besides texts (e.g., visual information), and conduct better cooperation between ID-based and modality-based item representations. 
We will also explore more sophisticated personalized prompts and user instructions to better extract accurate and informative knowledge from PLM for a better user experience in recommendation. 
For the practical deployment of PLM-based pre-trained recommendation models, the efficiency issue should also be carefully considered.



\appendix

\section*{Ethical Considerations}

This work focuses on personalized multi-domain sequential recommendation via pre-trained models. Sequential recommendation needs to take user historical behaviors as inputs. For evaluation, we conduct experiments on classical public recommendation datasets, where data masking techniques are conducted to protect user privacy. In practical systems, the proposed pre-trained recommendation model should be strictly used in the corresponding system, and the data should only be used under the authorization of users for privacy consideration. The fairness and security issues should also be considered in practical pre-trained (recommendation) models. Some pioneer efforts have been conducted in such fields, which is not the central concern of this work.


\appendix

\printbibliography
\end{document}